\documentclass[debug]{rmaa}

\usepackage{longtable,booktabs}
\usepackage{soul}
\usepackage{paralist}

\usepackage{psfrag,color}

\usepackage[latin1]{inputenc}

\title{New catalog of distances to planetary nebulae based on \textit{Gaia} parallaxes and statistical distances}

\author{
  Diego Hern\'andez-Ju\'arez,\altaffilmark{1} 
  M\'onica Rodr\'iguez,\altaffilmark{2}
  and Miriam Pe\~na\altaffilmark{1}}

\altaffiltext{1}{Instituto de Astronom\'ia, Universidad Nacional Aut\'onoma de M\'exico, Apdo.\ Postal 70264, Ciudad de M\'exico, Mexico}

\altaffiltext{2}{Instituto Nacional de Astrof\'isica, \'Optica y Electr\'onica, Luis Enrique Erro 1, Tonantzintla 72840, Puebla, Mexico}

\shortauthor{Hern\'andez-Ju\'arez, Rodr\'iguez \& Pe\~na}
\shorttitle{New catalog of distances to PNe}

\fulladdresses{
\item .
\item .
}

\listofauthors{Diego Hern\'andez-Ju\'arez,  M\'onica Rodr\'iguez, and Miriam Pe\~na}
\indexauthor{Diego Hern\'andez-Ju\'arez}
\indexauthor{M\'onica Rodr\'iguez}
\indexauthor{and Miriam Pe\~na}

\abstract{We have developed a method to determine the most reliable distances for a large group of planetary nebulae. For this purpose, we analyze the distances obtained from \textit{Gaia} parallaxes and three determinations of statistical distances. The most reliable distance is derived for 2211 objects, and uncertainties for these distances are calculated in a homogeneous way. Using our most reliable distances, we compare the distributions of Galactic heights of hydrogen-poor and hydrogen-rich central stars of planetary nebulae. We find that [WR] central stars are closer to the Galactic plane than hydrogen-rich central stars and than other hydrogen-poor central stars. The latter have a similar distribution to hydrogen-rich central stars, which is significantly different from the one of [WR] central stars. This result disagrees with the proposed evolutionary sequence for hydrogen-poor central stars. }

\resumen{Desarrollamos un m\'etodo que determina las distancias m\'as fiables de un grupo amplio de nebulosas planetarias a partir de las distancias obtenidas con paralajes de \textit{Gaia} y tres determinaciones de distancias estad\'isticas. Calculamos las distancias m\'as fiables para 2211 objetos y les asignamos incertidumbres. Con estas distancias, comparamos las distribuciones de alturas sobre el plano gal\'actico de objetos con estrellas ricas y pobres en hidr\'ogeno. Encontramos que las nebulosas planetarias con estrellas [WR] est\'an m\'as cerca del plano gal\'actico que aquellas con estrellas ricas en hidr\'ogeno y con otras estrellas pobres en hidr\'ogeno. Estas \'ultimas se distribuyen de manera similar a los objetos con estrellas ricas en hidr\'ogeno, y de forma significativamente distinta que los objetos con estrellas [WR]. Esto est\'a en desacuerdo con la secuencia evolutiva propuesta para estrellas centrales pobres en hidr\'ogeno. }

\addkeyword{planetary nebulae: general }
\addkeyword{stars: distances}
\addkeyword{parallaxes}

\begin{document}
\maketitle

\section{Introduction}
\label{section:introduction}
In the study of planetary nebulae (PNe) and stellar evolution, it is necessary to have accurate distances to calculate parameters such as luminosity, gaseous mass, and others. Several methods can be used to obtain distances to PNe, and they can be grouped into two classes: individual and statistical methods. Individual methods provide direct estimates of the distances to PNe. Some examples of individual distance estimate methods are trigonometric parallaxes, spectroscopic distances, expansion parallaxes, or the extinction method \citep[see the review by][ for more methods and examples]{2022PASP..134b2001K}. On the other hand, statistical estimates rely on the assumption that PNe have certain properties in common, or that they fulfill some empirical relation between two parameters, from one of which the distance can be derived. Some examples are the \citet{1956AZh....33..222S} method, based on a constant ionized mass for all PNe, and the relation between radio continuum surface brightness and physical radius of the nebula, first explored by \citet{1995A&A...293..541V}.

Statistical estimates are usually considered less reliable than individual estimates, but this is not always true. In some cases, statistical estimates have errors comparable to those of the individual estimates \citep{1995ApJ...446..279B}. Besides, very different distances are sometimes obtained with different individual methods for some objects \citep{1993ApJ...410..239Z,2022RAA....22h5013A}. 

The only individual method that is model independent and that in principle can lead to small uncertainties for a large quantity of PNe heliocentric distances is the trigonometric parallax of \textit{Gaia}. \textit{Gaia} is a space probe launched by the European Space Agency at the end of 2013 \citep{hodgkin2013transient,2016A&A...595A...1G,2020yCat.1350....0G}, and has provided trigonometric parallaxes for billions of objects \citep{2021A&A...649A...2L}. This has been a milestone in distance estimation for the astronomical community. However, not all of these distances are entirely reliable. Some parallaxes are negative or have very large errors. In those cases, Bayesian statistics must be used to infer distances from parallaxes \citep{bailer2021estimating}. However, the Bayesian estimates can be highly dependent on the assumed prior. Besides, even when the parallaxes are positive and have small errors, they could result from spurious solutions \citep{fabricius2021gaia}. In addition, it can be complicated to identify the central star of a PN in the \textit{Gaia} database. \citet{chornay2021one} and \citet{Gonzalez2021} have developed methods to identify the central stars of PNe, but the methods are not perfect and there may be misidentifications.

Due to these problems, we decided to revise the distances obtained with parallaxes from the Early Data Release 3 (EDR3) of \textit{Gaia}, in order to decide when it is necessary to use  other distance estimates. The aim of this work is to determine the most reliable distance estimate for a large sample of PNe. We explore several sets of statistical distances and the distances derived from \textit{Gaia} EDR3, compare them with each other, and create a procedure to determine the most reliable distance for each PN and its uncertainty.

This paper is organized as follows. In section~\ref{section:Sta_dis} we comment on the catalogs of statistical distances we will be using, in section~\ref{section:gaia} we explore the \textit{Gaia} parallaxes and their problems, and in section~\ref{section:comp} we compare the statistical distance estimates with the distances implied by the parallaxes. In section~\ref{section:proced} we present the procedure we follow to determine the most reliable distance estimate for each object, and in section~\ref{section:Compar_indiv} we compare our most reliable estimates with individual distance estimates available for several dozens of objects. In section~\ref{section:result} we use our most reliable distances to analyze the Galactic height distribution of hydrogen-poor and hydrogen-rich PNe. Finally, we present our conclusions in section~\ref{section:conclusion}.

\section{Statistical distances}
\label{section:Sta_dis}
We use the catalogs of statistical distances of  \citet{zhang1995statistical},  \citet{stanghellini2018galactic} and  \citet{frew2016halpha}. We choose these catalogs because they have the largest number of objects and are based on a variety of methods to calculate the distance, although some of them use the same observational data. The methods are described below.

 \citet{zhang1995statistical} uses two methods to estimate the distance. These methods are based on new calibrations of previously known empirical relations between the ionized mass and the brightness temperature with the intrinsic radius of PNe. The ionized mass and the brightness temperature are obtained from the flux at 5 GHz. To calibrate both methods,  \citet{zhang1995statistical} uses  a sample of 134 Galactic PNe with known individual distances determined by \citet{1993ApJ...410..239Z}. He provides the mean of the distances implied by both methods as his best estimate. His final sample contains 647 PNe.

The distances listed by \citet{stanghellini2018galactic} are based on the approach of \citet{stanghellini2008magellanic}, who recalibrate the method of \citet{1982ApJ...260..612D} using 70 Magellanic Cloud PNe. This method assumes that density-bounded PNe, which are optically thin to Lyman continuum radiation, have the same ionized mass, whereas both optically thick radiation-bounded PNe and bipolar PNe show a relation between their ionized mass and their surface brightness. Like \citet{zhang1995statistical}, \citet{stanghellini2008magellanic} calculate all their parameters using the PN angular sizes and the fluxes observed at 5 GHz. The final sample of \citet{stanghellini2018galactic} contains distances for 900 PNe.

 \citet{frew2016halpha} use an empirical relation between the H$\alpha$ surface brightness and the intrinsic radius of PNe to estimate their distances. They calibrate this relation using data for 322 PNe, of which 206 are Galactic and 126 are extragalactic objects. They find that optically-thick and optically-thin PNe have somewhat different behaviors and provide three relations, one for the full sample, one for optically thick objects, and one for optically thin PNe. They obtain distances for 1133 PNe, and for 515 of them they have the information required to estimate their optical thickness so that they can assign to these objects better distance estimates based on the relations for either optically thick or optically thin nebulae.

\subsection{Final sample of statistical distances}
Since \citet{frew2016halpha} and \citet{zhang1995statistical} have more than one distance estimate for each PN, we must decide which one to use. In the case of the distances of \citet{frew2016halpha}, we use those obtained with the formulas for optically thick and thin nebulae whenever possible (515 objects). Otherwise, we use the distance obtained with the general formula (618 objects).

We have compared the three distance estimates of \citet{zhang1995statistical}, the ones based on the ionized mass and the brightness temperature, and the mean of these two estimates, with the distances of \citet{stanghellini2018galactic} and \citet{frew2016halpha}. We find that the distances based on the brightness temperature method are in much better agreement with those of the other authors. For the other two sets of distances, 20$-$40\% of the objects have distances that disagree by more than 75\% from the distances of \citet{stanghellini2018galactic} and \citet{frew2016halpha}, whereas the brightness temperature method leads to this kind of disagreement for only 5 to 8\% of the PNe. Hence, we only use here the distances of \citet{zhang1995statistical} that are based on the brightness temperature method.

\section{Distances from \textit{Gaia} parallaxes}
\label{section:gaia}
In principle, the distances derived from \textit{Gaia} parallaxes, $p$, will be reliable when the objects are well identified and when the parallaxes are positive, do not have a considerable error, and are corrected for systematic errors. Besides, we must consider the quality of the fit to the astrometric observations. The RUWE (Renormalised Unit Weight Error) parameter is used to measure this quality \citep{2018A&A...616A...2L}. RUWE values above 1.4 suggest that there are problems with the astrometric solutions\footnote{\url{https://gea.esac.esa.int/archive/documentation/GDR2/Gaia_archive/chap_datamodel/sec_dm_main_tables/ssec_dm_ruwe.html}}, and we will not use those parallaxes here.

Identifying the central stars of PNe can be complicated, as the stars are faint, they can be hidden behind nebular material, and there might be several candidates in the central region of the nebula. Besides, some of the candidates may not be real objects; they can be \textit{Gaia} misidentifications arising from the effects of the surrounding gas and its nebular emission. \citet{chornay2021one} and \citet{Gonzalez2021} have developed methods to identify the central stars of PNe in  \textit{Gaia} EDR3. Both methods follow similar procedures: they look for the objects closest to the geometric center of the PNe and refine the selection by using colors, with \citet{Gonzalez2021} giving more importance to the latter criterion.

The catalogs by \citet{chornay2021one} and \citet{Gonzalez2021} contain in total 1140 objects with positive parallax and RUWE lower than 1.4, and they have in common 872 objects. From this sample in common, 25 objects have different identifications in the two catalogs, and eight of these 25 objects have estimates of statistical distances. If we use the identifications of \citet{chornay2021one} to determine the distances implied by the parallaxes (using the procedure described below), the differences between these distances and the statistical distances have an average lower than 0.05~dex, with a maximum difference of 0.5~dex. When the same procedure is done using the identifications of \citet{Gonzalez2021}, an average difference of more than 0.3~dex is obtained, with a minimum difference of 0.1~dex and a maximum of 0.7~dex. Therefore, we decided to use the identifications by \citet{chornay2021one} for the 872 objects in common, but we will also use the unique identifications of \citet{chornay2021one} for 190 objects, and those of \citet{Gonzalez2021} for 78 objects.

Once the objects are identified, the next step is to correct the parallaxes for systematic errors. We applied these \textit{zero-point} corrections using an available Python code that requires information on the source magnitudes, colors, and celestial positions to interpolate the values of the corrections \citep[][]{2021A&A...649A...4L}. 

Some objects in the \textit{Gaia} database have negative parallaxes or considerable errors. To be able to use the information for these objects, it is necessary to use a Bayesian approach, as \citet{bailer2021estimating} do. These authors calculate a distance estimate for each object with parallax in \textit{Gaia} EDR3 using Bayesian statistics. \citet{bailer2021estimating} use a prior based on a Milky Way model from the mock stellar catalog of \citet{Rybizki2020}, which is based on a three-dimensional model of the Galaxy. These distances should not be used indiscriminately, since for objects with large parallax errors, the distances converge to the prior \citep{2022MNRAS.516L..61O}. In this work, we will not use the results derived from negative parallaxes, and the distances based on  parallaxes with large errors will be used with caution, as described below.

There are problems in \textit{Gaia} parallaxes that cannot be completely solved, such as the spurious parallax solutions. \citet{fabricius2021gaia} mention that even in the region of parallaxes with errors smaller than 20\%, spurious parallax solutions can exist. To arrive at this result, they look for those objects with $p/\delta p<-5$, where they are sure that \textit{Gaia} is giving wrong results. \citet{fabricius2021gaia} find that at least 1.6\% of the objects have spurious solutions in this region, and consider that the same percentage will be present in the region of positive parallaxes with errors lower than 20\%.

In the next section, we explore further the problems that can arise from the use of \textit{Gaia} parallaxes in order to decide in which cases we should not use these data.

\section{Comparison of statistical and \textit{Gaia} distances}
\label{section:comp}

\begin{figure}   
  \begin{changemargin}{-2cm}{-2cm}
     \includegraphics[width=1.3\columnwidth]{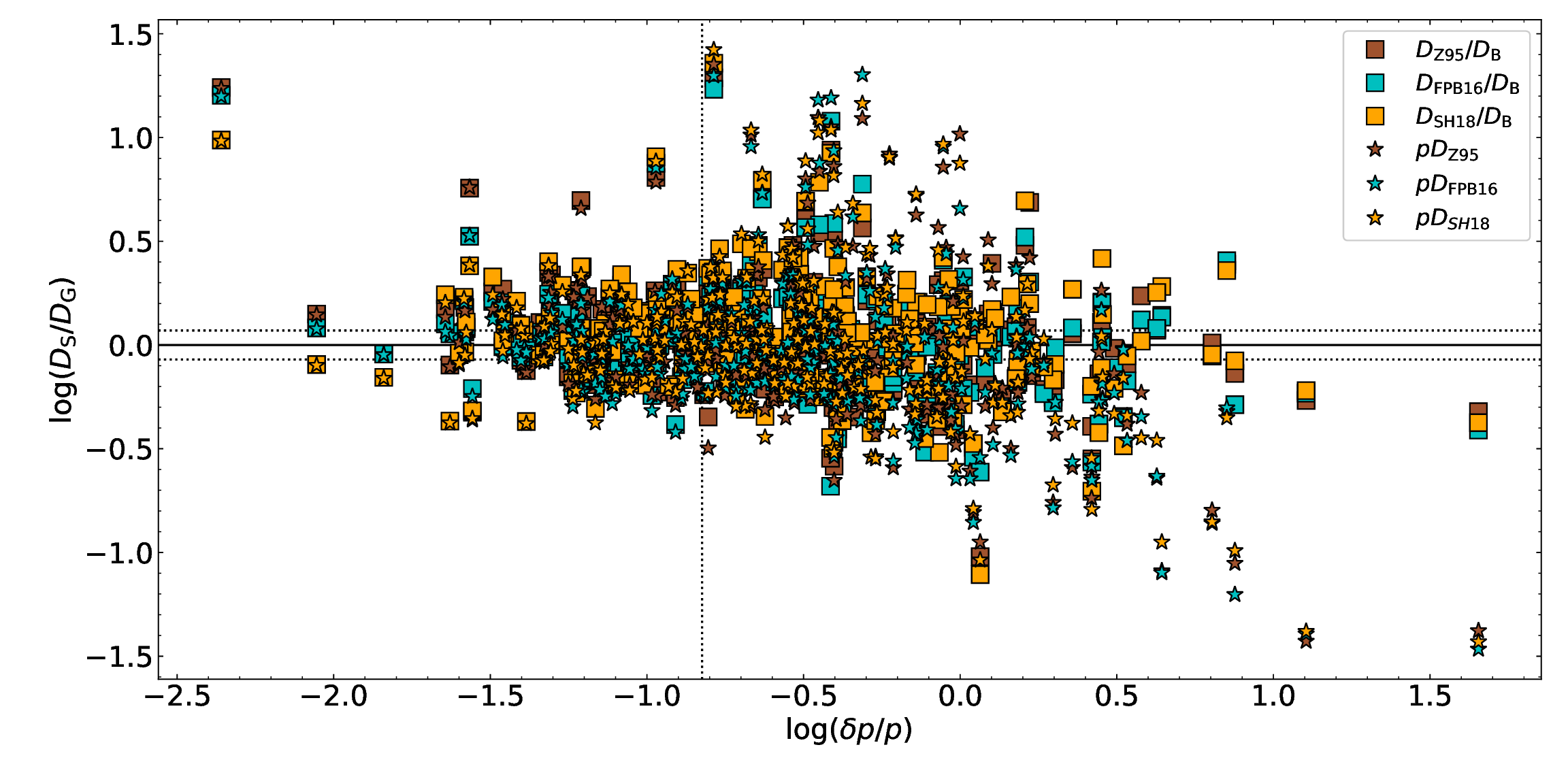}
     \caption{Ratio of the statistical distances, ${D_{\mathrm{S}}}$, and the distances derived from \textit{Gaia} parallaxes, $D_{\mathrm{G}}$, as a function of the relative error in the parallax, $\delta p/p$, for PNe that have statistical distances in the three catalogs that we are considering here: \citet[][blue symbols, $D_{\mathrm{FPB16}}$]{frew2016halpha}, \citet[][orange symbols, $D_{\mathrm{SH18}}$]{stanghellini2018galactic}, and \citet[][brown symbols, $D_{\mathrm{Z95}}$]{zhang1995statistical}. Stars are used for the comparison with distances derived from the inverse of \textit{Gaia} parallaxes and squares for the comparison with the Bayesian estimates ${D_{\mathrm{B}}}$. The vertical dotted line indicates a parallax error of 15\%, and the horizontal dotted lines correspond to differences of $\pm0.07$~dex.}
     \label{fig:comp1}
  \end{changemargin}
\end{figure}

We have selected a sample of 411 PNe that have positive parallaxes in \textit{Gaia} and the three estimates of statistical distances discussed above. For these objects, we are going to compare the two distances implied by the \textit{Gaia} parallax ($D_{\mathrm{G}}$) ---the one implied by the inverse of parallax, ${1}/{p}$, and the Bayesian estimate, $D_{\mathrm{B}}$--- with the statistical estimates (${D_{\mathrm{S}}}$) by \citet{frew2016halpha}, $D_{\mathrm{FPB16}}$, \citet{stanghellini2018galactic}, $D_{\mathrm{SH18}}$, and \citet{zhang1995statistical}, $D_{\mathrm{Z95}}$. This comparison is presented in Fig.~\ref{fig:comp1}, where we plot the ratio $D_{\mathrm{S}}/D_{\mathrm{G}}$ as a function of the relative error of the parallax, $\delta p/p$. We use stars for the distances derived from the inverse of the parallax and squares for the Bayesian distances. The statistical catalogs are identified with colors in the online version: blue for the distances of \citet{frew2016halpha}, orange for those of \citet{stanghellini2018galactic}, and brown for the distances of \citet{zhang1995statistical}.

We can see in Fig.~\ref{fig:comp1} that the three statistical distance estimates generally agree with each other. In fact, if we compare every possible pair of statistical distances (1744 pairs for 788 PNe), most of the differences (68\%, the traditional 1-$\sigma$ result) are lower than $\sim0.07$~dex, and 85\% of them are lower than $0.2$~dex. If we now compare the statistical values with those implied by \textit{Gaia}, we  see in Fig.~\ref{fig:comp1} that when the parallax errors are smaller than 15\% (the vertical dotted line in Fig.~\ref{fig:comp1}), $D_{\mathrm{G}}$ and $D_{\mathrm{S}}$ show a broad agreement for most of the objects, whereas for errors larger than 15\%, the differences between $D_{\mathrm{G}}$ and $D_{\mathrm{S}}$ are increasingly larger. On the other hand, for parallax errors $\sim$15\%, most of the  differences (68\%) between $1/p$ and $D_{\mathrm{B}}$ are smaller than 0.07~dex, and all of them are smaller than 0.2~dex, but, as Fig.~\ref{fig:comp1} shows, $1/p$ is an increasingly unreliable distance estimate as the parallax errors increase.

There are some objects in Fig.~\ref{fig:comp1} that show large differences between $D_{\mathrm{G}}$ and $D_{\mathrm{S}}$, even in the region of small parallax errors. An example is Abell~19, the object with the lowest parallax error in the figure. This is likely a misidentification of the PN central star. In fact,  \citet{Chornay2020} argue that the centrally located star is probably a nearby field star. Other objects with large differences are likely to be also misidentifications or to have spurious parallaxes.

Therefore, we will use 15\% as the defining line between the regions where $1/p$ is a good estimate of distance ($\delta p/p \le 0.15$) and where the Bayesian estimates are a better choice ($\delta p/p > 0.15$). Besides, we will only use the parallax distances when they show agreement (better than or equal to 0.07~dex, see \S \ref{section:proced}) with any of the statistical distance estimates, or when they are the only available distance estimates. Our approach is described below.

\section{Procedure for determining the most reliable distance for each PN and assignment of uncertainties}
\label{section:proced}

In order to obtain an extended catalog with as many distances as possible, we have compiled a sample of PNe that have at least one of the distance estimates considered here: the statistical distances of  \citet{frew2016halpha}, \citet{stanghellini2018galactic}, and  \citet{zhang1995statistical}, and the distances obtained from \textit{Gaia}.
In those cases where the PNe have several distance estimates and include the \textit{Gaia} parallax, we consider that the inverse of the parallax will provide the best distance estimate when the parallax errors are small, below 15\%, and when there is a good agreement between this distance and any of the statistical distance estimates. If this is not the case, we will use the median of the available values as the best distance estimate. When the parallax errors are larger than 15\%, we will only use the Bayesian estimate in our calculations. The procedure and the criteria we use are defined below.

\subsection{Inverse of the parallax}
\label{subsection:parallax}

As discussed in Section 4, the distance obtained from the inverse of the parallax is very similar to the Bayesian estimate when the parallax error is small, ${\delta p}/{p}<0.15$. We use in this case the inverse of the parallax as the \textit{Gaia} distance estimate. However, this distance must be similar to the statistical estimates in order to avoid problems with misidentifications and spurious parallaxes. Hence, in order to assign this distance to a given object, we require it to fulfill the conditions: ${\delta p}/{p}<0.15$ and
$\log(pD_{\mathrm{S}})\leq0.07$
for at least one statistical distance. There are 89 PNe that satisfy these requirements and they are classified as case~A in what follows.

In order to assign uncertainties to those distances, we explored three different approaches: we studied the uncertainties implied by the parallax errors; those implied by the Bayesian method \citep{bailer2021estimating}; and those obtained by comparing the inverse of the parallax with the second-farthest statistical distance to the inverse of parallax. The first two approaches lead to uncertainties lower than 10\% for 68\% of the objects, whereas for the last approach the 68th percentile is equal to 40\%. We decided to use the more conservative approach and assign this last uncertainty to these distances. We have decided to use this  approach because, even if the inverse of parallax shows some agreement with one statistical distance, the \textit{Gaia} parallaxes might still be affected by all the problems discussed in \S\ref{section:gaia}.
\subsection{Median of the available distances}

We will use the median of the available distances when there is more than one distance estimate, but no parallax is available, or the parallax error is too large, or the inverse of the parallax implies a distance that is very different from all the statistical distances.

For the case of large parallax errors, we have to decide whether to include the Bayesian estimate $D_{\mathrm{B}}$ in our calculations. In order to use this estimate, we require it to fulfill the same condition imposed on the inverse of the parallax in Section 5.1: $\log(D_{\mathrm{S}}/D_{\mathrm{B}})\leq0.07$
for at least one statistical distance. If this condition is met, our most reliable distance is given by the median of the statistical values and the Bayesian value (177 objects, case~B in what follows). If not, our most reliable distance is given by the median of the statistical values (635 objects, case~C in what follows).  In order to minimize the effect of very anomalous distances on the final results, the median is calculated for the logarithmic distances.

We explored two possible ways to assign uncertainties to the median values: the Bayesian uncertainty and the uncertainty obtained by comparing the median of the distances with the most extreme distance. Both distributions are very similar, and the 68th percentile is at $\sim40$\%. Hence we assign 40\% uncertainties to these distance estimates. We do not use the Bayesian uncertainties because our estimate is not necessarily based on the Bayesian distance and this distance might still be affected by all the problems discussed in \S\ref{section:gaia}.

\subsection{PNe with only one distance estimate}
  
 Some PNe have a single distance estimate, either obtained from the \textit{Gaia} parallax (the inverse of the parallax for errors below 15\% and the Bayesian estimate for larger errors) or from one statistical method. Besides, some PNe have just one statistical distance estimate and it disagrees with the results implied by the \textit{Gaia} parallax, and this implies that we only use the one statistical value to assign the distance to these objects. In total, there are 1310 PNe whose distances are based on a single distance estimate, which are labeled as case~D in what follows.
 
 In order to provide uncertainties for these distances, we explored the uncertainties assigned by \citet{bailer2021estimating} to their Bayesian estimates for those PNe where we use this distance estimate. These uncertainties are lower than $\sim60$\%. We also studied the distribution of distance differences for PNe that have only two distance estimates. We find that 68\% of these differences are lower than 60\%. Since the results for PNe with only one distance estimate can be even less reliable, we decided to assign them a larger error. Hence, if the object has a single distance estimate, we assign a relative error of 70\%.
 
\subsection{Final results}

In Table~A in Appendix \ref{sec:table-distances} we present our final results for 2211 PNe. The table lists the distances provided in the three statistical catalogs we are using: \citet[][$D_{\mathrm{FPB16}}$]{frew2016halpha}, \citet[][$D_{\mathrm{SH18}}$]{stanghellini2018galactic}, and \citet[][$D_{\mathrm{Z95}}$]{zhang1995statistical}. Besides, Table~A also shows the Bayesian distance of \citet[][$D_{\mathrm{B}}$]{bailer2021estimating}, the distance implied by the \textit{Gaia} parallax and its uncertainty, our final distance estimate, $D_{\mathrm{tw}}$, with its uncertainty, and the method used to determine this final estimate. The complete table can be found in the online version.

In the final results, we use some distance from \textit{Gaia} for 959 objects (the inverse of the parallax or the Bayesian distance). We use the inverse of the parallax because it agrees, to within $\pm0.07$~dex, with some statistical distance in 89 cases. We use the median of statistical distances and Bayesian distances (because the latter agrees, to within $\pm0.07$~dex, with some statistical distance) in 177 cases. In 693 cases, the \textit{Gaia} estimate is the only available distance estimate.  Finally, in 1075 objects, the distances are based on one statistical distance or the median of the available statistical distances. Thus, the \textit{Gaia} data are used in 43\% of the cases and in some objects we only use statistical estimates even though the \textit{Gaia} parallax is available. 

In summary, we use the \textit{Gaia} distance for 959 PNe, statistical distances for 1075 PNe, and the median of one Gaia distance and the statistical distances for 177 objects. In total, we have distances for 2211 PNe.

\section{Comparison with individual distance measurements}
\label{section:Compar_indiv}

In this section, we compare our most reliable distances (those that are based on more than one distance estimate) with some individual distance estimates. We use the individual distance estimates for 48 PNe of the calibration sample of \citet{frew2016halpha}; 9 PNe from \citet{Yang2016}; 2 PNe from \citet{schonberner2018}; 2 PNe from  \citet{GomezGordillo2020}; and 12 PNe from  \citet{Dharmawardena2021}. The results of \citet{Dharmawardena2021} are extinction distances based on optical and radio data. For some objects they provide two estimates of the extinction and these can be very different. Hence, we only use their results for PNe that have two estimates of the extinction that differ in less than 60\%. The individual distance estimates that we are using are based on a variety of methods and we select only those distances that have uncertainties lower than 25\%. Some objects have two distance estimates and in this case we use the mean value. The final comparison sample has 65 PNe.

\begin{figure}
	\includegraphics[width=1\columnwidth]{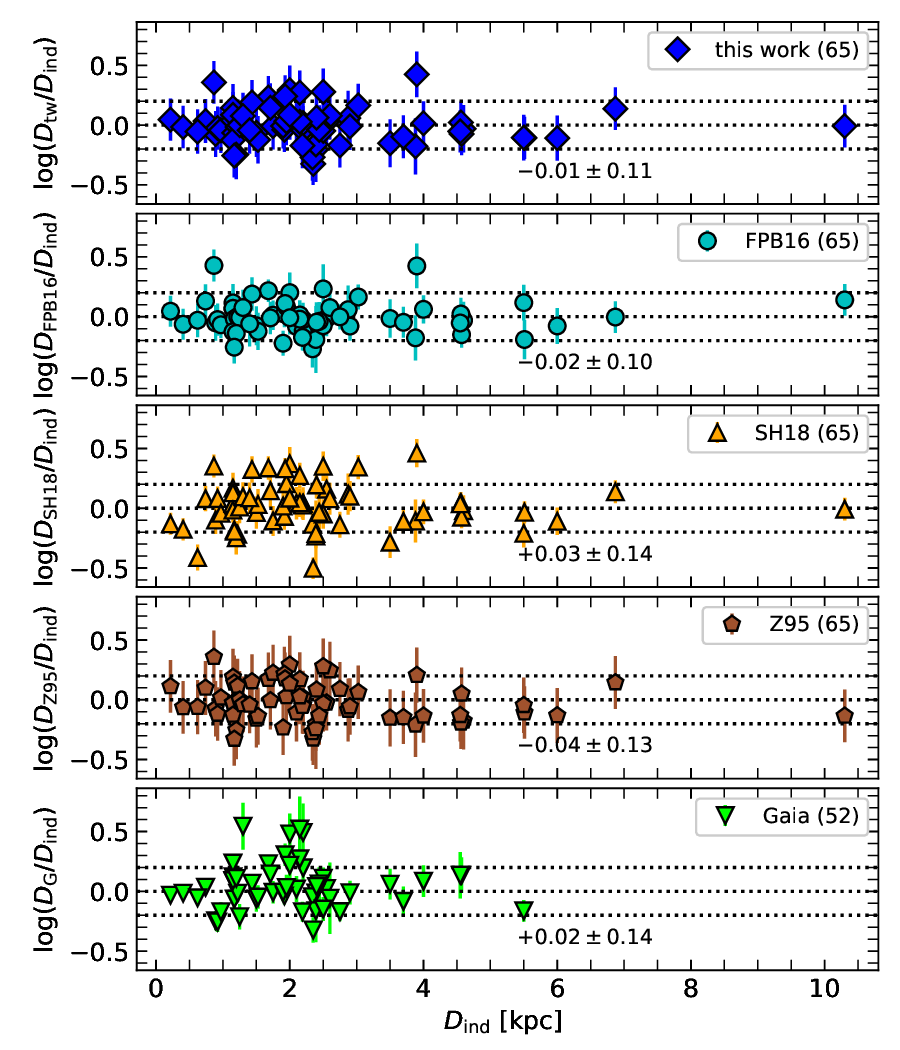}
	\caption{Ratio of the distances derived in this work ($D_{\mathrm{tw}}$, top panel) and the statistical distances of \citet[][$D_{\mathrm{FPB16}}$, second panel]{frew2016halpha}, \citet[][$D_{\mathrm{SH18}}$, third panel]{stanghellini2018galactic}, and \citet[][$D_{\mathrm{Z95}}$, fourth panel]{zhang1995statistical} to some individual distance estimates, $D_{\mathrm{ind}}$, as a function of $D_{\mathrm{ind}}$ for 65 PNe. The same comparison is performed for 52 of these PNe that have positive \textit{Gaia} parallaxes and RUWE values lower than 1.4 (bottom panel). The horizontal dotted lines correspond to differences of zero and 60\%. The median and the mean absolute deviation of the plotted values are shown in each panel.}
	\label{fig:comIndiv}
\end{figure}

In Fig.~\ref{fig:comIndiv}, we show the comparison between our final distances, $D_{\mathrm{tw}}$, and the individual distances, $D_{\mathrm{ind}}$, as a function of $D_{\mathrm{ind}}$. We also show this comparison for the three statistical distance estimates that we are using: $D_{\mathrm{FPB16}}$ \citep{frew2016halpha}\footnote{Please note that in this case we are comparing the individual distances used by \citet{frew2016halpha} to calibrate their statistical method with the statistical distances they obtain for the same objects.}, $D_{\mathrm{SH18}}$ \citep{stanghellini2018galactic}, and $D_{\mathrm{Z95}}$ \citep{zhang1995statistical}. Finally, the bottom panel compares the distances implied by the \textit{Gaia} parallaxes (the inverse of the parallax for errors lower than 15\% and the Bayesian estimate for larger errors) with the individual estimates for the 52 objects that had this information. All these parallaxes are positive and have RUWE values lower than 1.4. Three horizontal dotted lines at zero and 60\% differences are plotted in Fig.~\ref{fig:comIndiv} for reference. Besides, the values of the median and mean absolute deviation of the plotted results are shown in each panel of Fig.~\ref{fig:comIndiv}.

We can see in Fig.~\ref{fig:comIndiv} that our final distances, the three statistical distance estimates, and the results from \textit{Gaia} broadly agree with the individual estimates for this sample of PNe, with our results and those of \citet{frew2016halpha} showing the best agreement. In fact, 68\% of our results have agreements better than 40\%, the uncertainty that we have assigned to the distances plotted in the top panel of Fig.~\ref{fig:comIndiv}.

\section{Distances to the galactic plane for H-rich and H-poor central stars of PNe}
\label{section:result}

The spectral types of central stars of planetary nebulae (CSPNe) can give us information about the processes that they have undergone and about their evolutionary state. There are several types of CSPNe, and most of them can be grouped into two major groups: hydrogen-rich (HR) and hydrogen-poor (HP) CSPNe \citep{Mendez1991}. 

The origin of HP CSPNe is not entirely clear. It has been proposed that during their post-AGB phase, some stars experience a very late thermal pulse, which returns them to the AGB phase \citep{Iben1984}, where the hydrogen is hidden or lost by winds.  CSPNe that go through this event are known as \textit{born-again} stars. This is the most studied origin, but so far only eight HP CSPNe have been proven to be \textit{born-again} \citep{jacoby2020}.  \citet{Fang2014} and  \citet{gorny2000} mention that most CSPNe are unlikely to have a very late thermal pulse. Therefore, this phenomenon  does not seem responsible for the known HP CSPNe which amount to about 30\% of the total CSPNe that have a well-defined spectral type  \citep{weidmann2020catalogue}.

Other possible scenario to explain the existence of HP stars is that they  descend from a close  binary system \citep[see, e.g.,][]{Tylenda1993, Orsola2003RMx}. However, in many cases  it is difficult to prove that the central star is part of a binary system, and as a result, it is not clear how many CSPNe could be converted to HP CSPNe in this way.

The origin of HP CSPNe could be related to the mass of the progenitor, either because a massive progenitor could solve some problems with the \textit{born-again} phenomenon \citep{AckerGorny1996A&A...305..944A} or because of some undefined phenomena related to other possible scenarios. \citet{{1982IAUS...99..423H}} propose that [WR] CSPNe, which are HP, come from a massive progenitor because they have higher luminosities than other types. This hypothesis has been explored using chemical abundances, in particular the N/O ratio. \citet{1995A&A...303..893G} find no evidence that [WR] CSPNe come from massive progenitors, but \citet{2013A&A...558A.122G} estimate that about half of their sample of [WC] CSPNe had initial masses larger than 4 M$_\odot$. However, determining the mass of the progenitor star using chemical abundances can be quite complicated due to the uncertainties involved in calculating abundances. 

A better option is to study the distances to the Galactic plane, which should be lower for massive progenitors. Some comparisons of the Galactic distributions of [WR] CSPNe (and other HP CSPNe) and HR CSPNe have been made. For example, \citet{AckerGorny1996A&A...305..944A} compare the distribution of Galactic latitudes of 47 [WR] CSPNe with those of their total sample of 350 HR and HP ([WR] CSPNe included) CSPNe, and do not find significant differences. On the other hand, \citet{weidmann2011} and \citet{weidmann2020catalogue} use larger samples (397 and 443 CSPNe, respectively) to compare the distributions of Galactic latitude of HP (including [WR]) and HR CSPNe.  \citet{weidmann2011} have 205 HP objects (106 of them [WR]) and \citet{weidmann2020catalogue} have 153 HP CSPNe (with 123 [WR]). Both works conclude that HP CSPNe are found at lower Galactic latitudes than HR CSPNe.

The different results found by the three studies might be due to the different sample sizes or could arise from the fact that \citet{AckerGorny1996A&A...305..944A} make their comparison between [WR] and all CSPNe, and not between two different groups (such as HR and [WR] CSPNe). Nevertheless, a problem shared by the three works is that they do not consider the distributions of heliocentric distances of their objects. \citet{gorny2004populations} find that [WR] CSPNe are concentrated towards the Galactic center. This means that the different Galactic distributions of [WR] and HR objects could lead to differences in their distributions of Galactic latitudes.

\citet{Pena2013} compare distances to the Galactic plane of [WR] CSPNe and HR CSPNe using a total sample of 77 CSPNe (46 of them [WR]). They note that [WR] CSPNe are more concentrated towards the Galactic plane than HR CSPNe. Although \citet{Pena2013} consider distances, their sample is small.

We study here the distribution of distances to the Galactic plane ($z_\mathrm{G}$) using our distance catalog and the spectral types compiled by \citet{weidmann2020catalogue}. We only use CSPNe of this catalog that have well-defined spectral types and that can be clearly classified as HP or HR. We calculate $z_\mathrm{G}$ for 81 HP CSPNe and 187 HR CSPNe. In the HP CSPNe group, we distinguish between [WR] CSPNe and non-[WR] CSPNe, such as PG\,1159 and DO white dwarfs. 

\begin{figure}
	\includegraphics[width=\columnwidth]{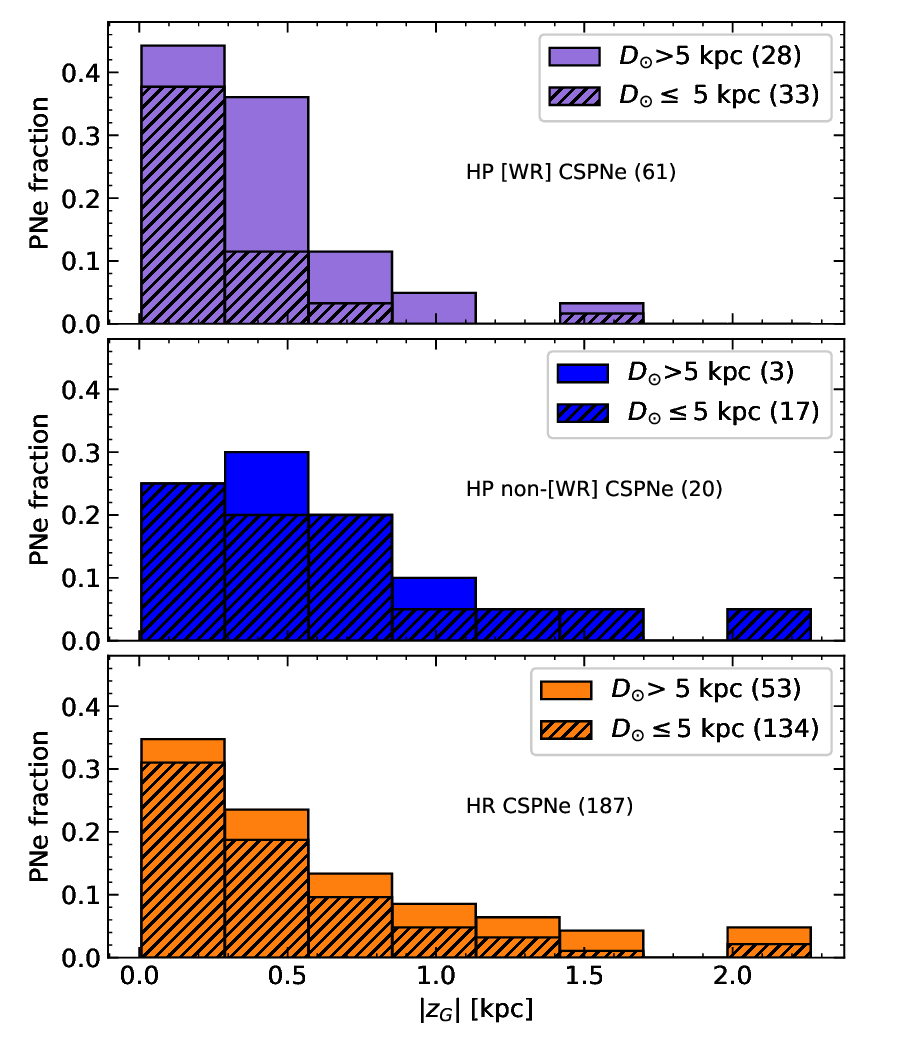}
    \caption{Distributions of distances to the galactic plane ($z_\mathrm{G}$) for [WR] CSPNe (top panel), HP non-[WR] CSPNe (middle panel) and HR CSPNe (bottom panel) for objects at any heliocentric distance and for objects with heliocentric distances lower than 5 kpc (hatched areas). The number of objects in each category are shown within parentheses.}
    \label{fig:Z_G_WR}
\end{figure}

Fig.~\ref{fig:Z_G_WR} shows the resulting distributions of $z_\mathrm{G}$ of [WR], HP non-[WR] and HR CSPNe. We can see in this figure that [WR] CSPNe are indeed closer to the Galactic plane than HR CSPNe, as previously found by \citet[][]{Pena2013}. A Kolmogorov-Smirnov test \citep{press2007numerical} shows that the $z_\mathrm{G}$ distributions of [WR] CSPNe and HR CSPNe are significantly different, with a $p$-value of 0.008 (the probability of obtaining differences equal or greater than those observed if both distributions come from the same parent distribution). The $z_\mathrm{G}$ distribution of HP non-[WR] is also different from that of [WR] CSPNe ($p$-value equal to 0.01), but compatible with the distribution of HR CSPNe ($p$-value of 0.2). 

However, most of our non-[WR] HP objects are close to the Sun, probably because of their lower brightness. Besides, as found by \citet{gorny2004populations}, our [WR] CSPNe are more abundant towards the Galactic center. Therefore, we restricted our sample to objects with heliocentric distances lower than 5 kpc in order to avoid as much as possible introducing biases in the distributions of $z_\mathrm{G}$. The [WR] CSPNe closer to the Galactic center are excluded with this condition, but we still retain an acceptable amount of HP non-[WR] CSPNe. These objects are shown in Fig. \ref{fig:Z_G_WR} with the hatched areas. The $z_\mathrm{G}$ for these PNe are shown in Table~B in Appendix \ref{sec:tablez}. In the first column of this table the object is indicated, in the second column the group, in the third column the spectral type and in the last column,  $z_\mathrm{G}$ and its uncertainty.  The complete table is published in the online version.

In this restricted sample, there are 134 HR CSPNe, 17 HP non-[WR] CSPNe (including 12 PG\,1159, one O(He) and four WD) and 33 [WR] CSPNe. Like the massive Pop.\ I WR stars, the [WR] stars are classified into two groups \citep{1998MNRAS.296..367C}: the early [WC] stars (types [WC 4] and [WO 4-1]  and the late [WC] stars (types [WC 12-5]). In the sample with distances lower than 5 kpc, there are 6 late [WC] and 24 early [WC] CSPNe. 

A Kolmogorov-Smirnov test for the restricted sample shows that the difference in the $z_\mathrm{G}$ distributions of [WR] and HR CSPNe is even more significant ($p$-value of 0.0004); the differences between HP non-[WR] and [WR] CSPNe are still significant ($p$-value of 0.025), and the distributions of HP non-[WR] and HR CSPNe are also significantly different ($p$-value of 0.048) instead of compatible, as found with the full sample. The median values of $|z_\mathrm{G}|$ are 0.55 kpc for HP non-[WR] objects, 0.34 kpc for HR CSPNe, and 0.19 kpc for [WR] CSPNe.

 The different result obtained in the comparison between the distributions of $|z_\mathrm{G}|$ for HP non-[WR] and HR CSPNe for heliocentric distances below 5 kpc and for the full sample, is due to the effect of two HP non-[WR] with spectral type O(He) that have the largest Galactic heights in the HP non-[WR] group, $|z_\mathrm{G}|=1.6$, $2.0$ kpc. If we remove from this group the five non-PG\,1159 objects and apply the Kolmogorov-Smirnov test, we find that the distributions of PG\,1159 and HR CSPNe are again compatible ($p$-value of 0.98). On the other hand, the PG\,1159 objects still have a significantly different distribution of $|z_\mathrm{G}|$ from the [WR] CSPNe ($p$-value of 0.012).

As \citet{Pena2013} did, we find that [WR] CSPNe seem to be related to massive progenitors. However, this is not true for the other HP CSPNe, like the PG\,1159 objects. This result disagrees with the commonly assumed evolutionary sequence for HP CSPNe, in which late [WR] stars evolve to early [WR] stars and these, in turn, evolve to PG\,1159 stars \citep{AckerGorny1996A&A...305..944A,gorny2000,werner2006,weidmann2020catalogue}. It seems clear from our results that the [WR] CSPNe are not the progenitors of the other HP CSPNe. 

\section{Conclusions}
\label{section:conclusion}

We have developed a method to determine the most reliable distance for 2211 PNe using the distances derived from \textit{Gaia} data and three catalogs of statistical distances \citep{zhang1995statistical,frew2016halpha,  stanghellini2018galactic}. We show that the \textit{Gaia} distances are not always reliable for CSPNe and they must be used with caution. We also assign uncertainties to our final distances using a homogeneous approach. We find that our distances and those of \citet{frew2016halpha} show the best agreements with individual distance estimates.
With the next data releases from \textit{Gaia}, we can expect to increase the number of PNe with well determined distances and to improve the reliability of our catalog. 

We use our catalog to study the distributions of distances to the Galactic plane of HR and HP CSPNe. We find that there are differences between these distributions, arising from the lower heights above the Galactic plane of [WR] CSPNe. On the other hand, HP non-[WR] stars, especially the PG\,1159 objects, have a distribution similar to the one followed by HR CSPNe. These results suggest that [WR] CSPNe have more massive progenitors and that there does not seem to exist an evolutionary sequence from [WR] CSPNe to PG\,1159 stars, as commonly believed.
\section*{Acknowledgements}

We thank the anonymous referee for her/his very useful comments. This work received financial support of grant UNAM PAPIIT IN111423. DH-J acknowledges a scholarship from CONAHCYT, Mexico. This work has made use of data from the European Space Agency (ESA) mission
{\it Gaia} (\url{https://www.cosmos.esa.int/gaia}), processed by the {\it Gaia}
Data Processing and Analysis Consortium (DPAC,
\url{https://www.cosmos.esa.int/web/gaia/dpac/consortium}). Funding for the DPAC
has been provided by national institutions, in particular the institutions
participating in the {\it Gaia} Multilateral Agreement. 

\begin{appendices}
\section{Table of available distances and our chosen distances}\label{sec:table-distances}
\begin{changemargin}{-3cm}{-1cm}
\begin{longtable}{l c c c c c c l l}
\hline
PN G & $D_{\mathrm{Z95}}$ & $D_{\mathrm{FBP16}}$ & $D_{\mathrm{SH18}}$ & $D_{\mathrm{B}}$ & 1/$p$ & $\delta p$/$p$ & $D_{\mathrm{tw}}$ & Case \\
\hline \hline
\endfirsthead
\hline
PN G & $D_{\mathrm{Z95}}$ & $D_{\mathrm{FBP16}}$ & $D_{\mathrm{SH18}}$ & $D_{\mathrm{B}}$ & 1/$p$ & $\delta p$/$p$ & $D_{\mathrm{tw}}$ & Case \\
\hline \hline
\endhead
\hline
\multicolumn{9}{c}{Continue on the next page.}
\endfoot
\endlastfoot

000.0$-$01.0 & \nodata & \nodata & \nodata & 8.20 & 1.22 & 0.48 & $8.2\pm5.7$ & D\\
000.0$-$02.5 & \nodata & \nodata & \nodata & 6.73 & 1.82 & 1.62 & $6.7\pm4.7$ & D\\
000.0$-$06.8 & \nodata & 6.97 & \nodata & 8.39 & 21.28 & 0.72 & $7.0\pm4.9$ & D\\
000.1$+$02.6 & \nodata & \nodata & \nodata & 4.88 & 3.82 & 0.62 & $4.9\pm3.4$ & D\\
000.1$+$17.2 & 8.87 & 9.46 & 14.51 & 7.82 & 34.56 & 2.28 & $9.2\pm3.7$ & B\\
000.1$-$01.1 & 6.05 & \nodata & 8.63 & \nodata & \nodata & \nodata & $7.2\pm2.9$ & C\\
000.1$-$01.7 & \nodata & 6.95 & \nodata & \nodata & \nodata & \nodata & $7.0\pm4.9$ & D\\
000.1$-$02.3 & \nodata & 7.62 & 5.58 & \nodata & \nodata & \nodata & $6.5\pm2.6$ & C\\
\hline
\end{longtable}
\end{changemargin}
\newpage
\section{Table of distances to the Galactic plane for CSPNe with heliocentric distances lower than 5 kpc.}\label{sec:tablez}
\begin{changemargin}{3cm}{2cm}
\begin{longtable}{l l l r}
\hline
PN G & Type & Spectral Type & $z_{\rm G}~(kpc) $   \\
\hline \hline
\endfirsthead
\hline
PN G & Type & Spectral Type & $z_{\rm G}~ (kpc)$  \\
\hline \hline
\endhead

\hline
\multicolumn{4}{c}{Continue on the next page.}
\endfoot
\endlastfoot

002.2$-$09.4 &[WR]& [WO 4]pec & $-0.74\pm0.30$ \\
002.4$+$05.8 &[WR]& [WO 3] & $0.116\pm0.046$ \\
003.1$+$02.9 &[WR]& [WO 3] & $0.146\pm0.058$ \\
011.9$+$04.2 &[WR]& [WO 4]pec & $0.26\pm0.10$ \\
017.9$-$04.8 &[WR]& [WO 2] & $-0.40\pm0.16$ \\
020.9$-$01.1 &[WR]& [WO 4]pec & $-0.044\pm0.017$ \\
027.6$+$04.2 &[WR]& [WC 7-8] & $0.31\pm0.13$ \\
029.2$-$05.9 &[WR]& [WO 4] & $-0.26\pm0.10$ \\
\hline
\end{longtable}
\end{changemargin}

\end{appendices}
\end{document}